\documentclass[apjl]{emulateapj}
\usepackage[figuresright]{rotating}
\usepackage{subfigure}
\usepackage{epic,eepic}
\usepackage{graphicx}

\shorttitle{OPTICAL STUDIES ON M82-X1}
\shortauthors{Wang et al.}

\begin{document}

\title{TWO OPTICAL COUNTERPART CANDIDATES OF M82 X-1 FROM HST OBSERVATIONS}

\author{Song Wang\altaffilmark{1}, Jifeng Liu\altaffilmark{1}, Yu Bai\altaffilmark{1}, AND Jincheng Guo\altaffilmark{1,2}}

\altaffiltext{1}{Key Laboratory of Optical Astronomy, National Astronomical Observatories,
Chinese Academy of Sciences, Beijing 100012, China;
jfliu@bao.ac.cn, songw@bao.ac.cn}
\altaffiltext{2}{University of Chinese Academy of Sciences, Beijing 100049, China}

\begin{abstract}

Optical counterparts can provide significant constraints on the physical nature of
ultraluminous X-ray sources (ULXs). In this letter,
we identify six point sources in the error circle of a ULX in M82, namely M82 X-1,  by registering {\it Chandra} positions onto
{\it Hubble Space Telescope} images.
Two objects are considered as optical counterpart candidates of M82 X-1, which
show F658N flux excess compared to the optical continuum that may suggest the existence of an accretion disk.
The spectral energy distributions of the two candidates match well with the spectra
for supergiants, with stellar types as F5-G0 and B5-G0, respectively.
Deep spatially resolved spectroscopic follow-up and detailed studies are needed
to identify the true companion and confirm the properties of this BH system.

\end{abstract}


\section{INTRODUCTION}
\label{intro.sec}

Ultraluminous X-ray sources (ULXs) are extranuclear sources with
an observed luminosity in excess of 10$^{39}$ erg/s, which are possibly
the long-sought intermediate mass black holes (IMBHs), or
stellar mass black holes (SMBHs) in a new ultraluminous accretion state \citep{Gladstone2009}.
M82 X-1 is one of the most promising IMBH candidates based on multiple X-ray properties:
the extremely high luminosities \citep{Matsumoto2001, Kaaret2006b}, the quasi-periodic oscillation (QPO) behaviors
in the 0.05--0.1 Hz frequency range \citep{Strohmayer2003, Feng2007}, the new-identified thermally dominant states
\citep{Feng2010}, and a 3:2 QPO at frequencies of 3.32 and 5.07 Hz \citep{Pasham2014}, all of which point to an
IMBH of several hundred solar masses or above, although some models \citep{Okajima2006} suggested M82 X-1
may be a massive SMBH (19--32 $M_\odot$) shining at a super-Eddington luminosity.

The optical counterpart may provide crucial diagnostics about the nature of one ULX.
The magnitudes and colors of the optical counterpart may help reveal the type of the donor star and
distinguish between SMBHs and IMBHs \citep{Madhusudhan2008}, while optical spectroscopic
monitoring can be used to obtain dynamical mass estimates for the BH system \citep{Liu2013}.
Many efforts have been done to search for the optical counterpart of M82 X-1,
which was first reported to be associated with
a young massive star cluster MGG 11 \citep{Portegies Zwart2004}.
However, with more careful astrometry from {\it HST} Near Infrared Camera
and Multi-Object Spectrometer (NICMOS) and Wide Field Camera 3 (WFC3)/Infrared Channel images,
\citet{Kong2007} and \citet{Voss2011} argued that M82 X-1 is $\sim$ 0.65 arcsec away from MGG 11,
with an offset significance of 3$\sigma$.
Using archive images obtained by the Second Wide Field and Planetary Camera,
no optical counterpart was found for M82 X-1 \citep{Ptak2006}.
Recently, \citet{Gladstone2013} reported two candidate counterparts of M82 X-1 using the
Advanced Camera for Surveys (ACS)/Wide Field Camera (WFC)
observations, however, both of them are resolved into several sources in the images
of ACS/High Resolution Camera (HRC).
Motivated by these studies presented above, we decide to obtain
more precise location of the optical counterpart for follow-up studies on M82 X-1.

In this letter, we reexamine the {\it HST} data to perform accurate astrometry and photometry for M82 X-1.
Two candidate optical counterparts of M82 X-1 are identified, and some constraints on their physical nature are
discussed from the multiband spectral energy distribution (SED).
We describe our data analysis and results in Section \ref{data.sec},
and present SEDs for the counterparts in Section \ref{sed.sec}.
A discussion follows in Section \ref{discuss.sec}.

\section{DATA ANALYSIS AND RESULTS}
\label{data.sec}

\subsection{Astrometry}
\label{astro.sec}

For precise localization of M82 X-1 on optical images, we use multiple matched pairs of objects between {\it Chandra} and {\it HST} observations with the help of
Sloan Digital Sky Survey (SDSS) observations.
For the {\it Chandra} image, we select one of the deepest ACIS observations (Obsid 10543),
and used the CIAO tool {\sc WAVDETECT} to detect X-ray sources.
\citet{Xu2015} provided X-ray position for M82 X-1 (R.A.=9:55:50.123, decl.=+69:40:46.54) using the same
observation, which is adopted as the original {\it Chandra} position of the ULX.
The region around M82 X-1 has been observed frequently with {\it HST} using ACS, including HRC and
WFC, and WFC3/Ultraviolet-Visible Channel (UVIS).
Here the ACS/WFC F555W observation (Data Set j9l021d8q) is selected for the wide field of view ($\sim$ $202''\times202''$).
The observation from SDSS Data Release 12 is used for intermediate comparisons,
because it is hard to find matches that can be used for a direct registration of
{\it Chandra} and {\it HST} images \citep{Voss2011}.
The matched pairs of objects in the comparisons are presented in Table \ref{table1}.
The astrometric errors for the SDSS-{\it Chandra} and SDSS-{\it HST} comparisons are
$0.44''$ and $0.18''$ in right ascension, while $0.33''$ and $0.02''$ in declination.
The final correction used to translate the {\it Chandra} position of M82 X-1 onto the {\it HST} image is
$1.74''\pm0.48''$ in right ascension and $0.30''\pm0.33''$ in declination. The uncertainties are a quadratic
sum of the standard deviations of these comparisons.

\begin{figure*}[!htb]
\figurenum{1}
\center
\subfigure{\includegraphics[width=0.4\textwidth]{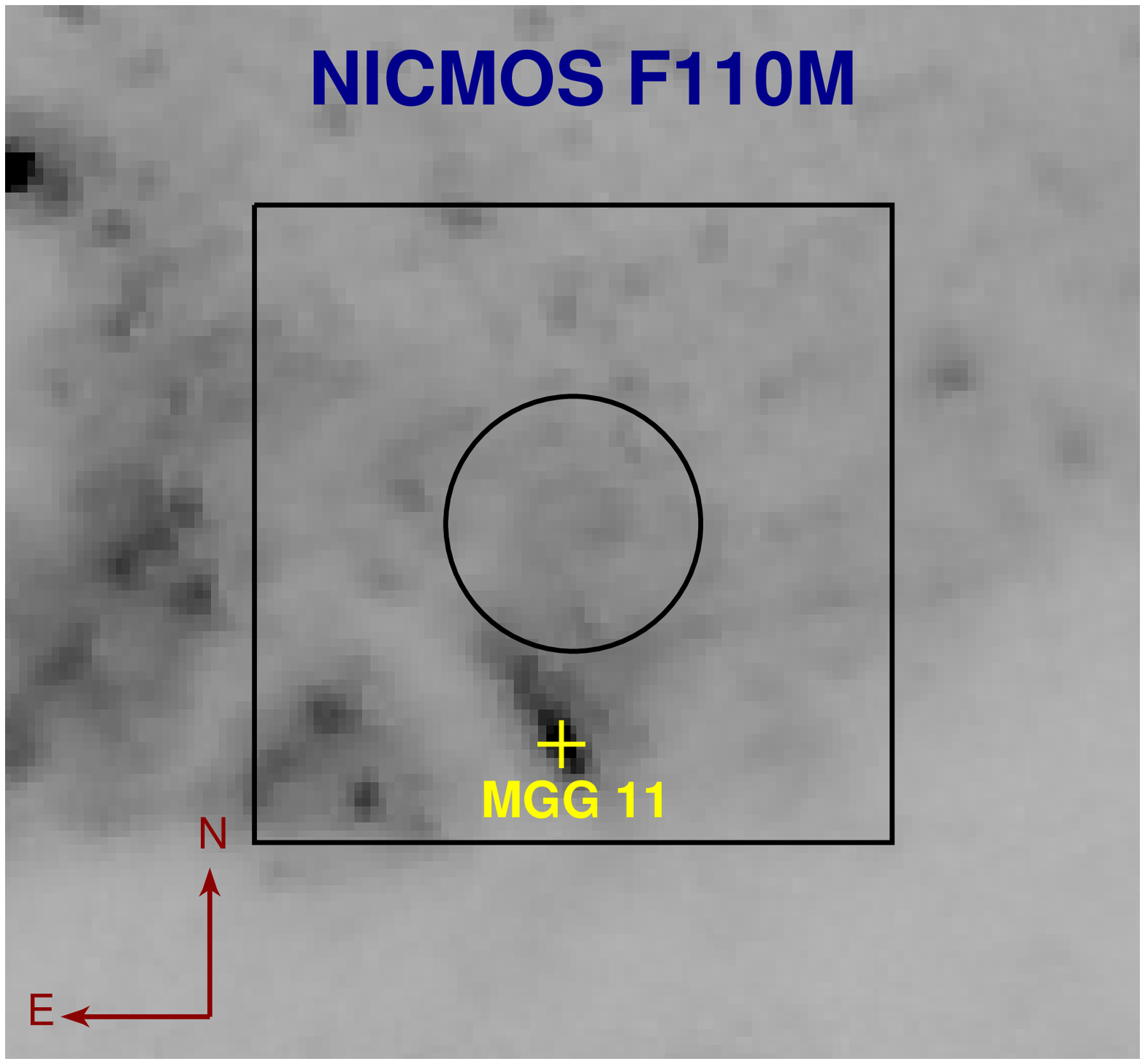}}
\hspace{2 mm}
\subfigure{\includegraphics[width=0.376\textwidth]{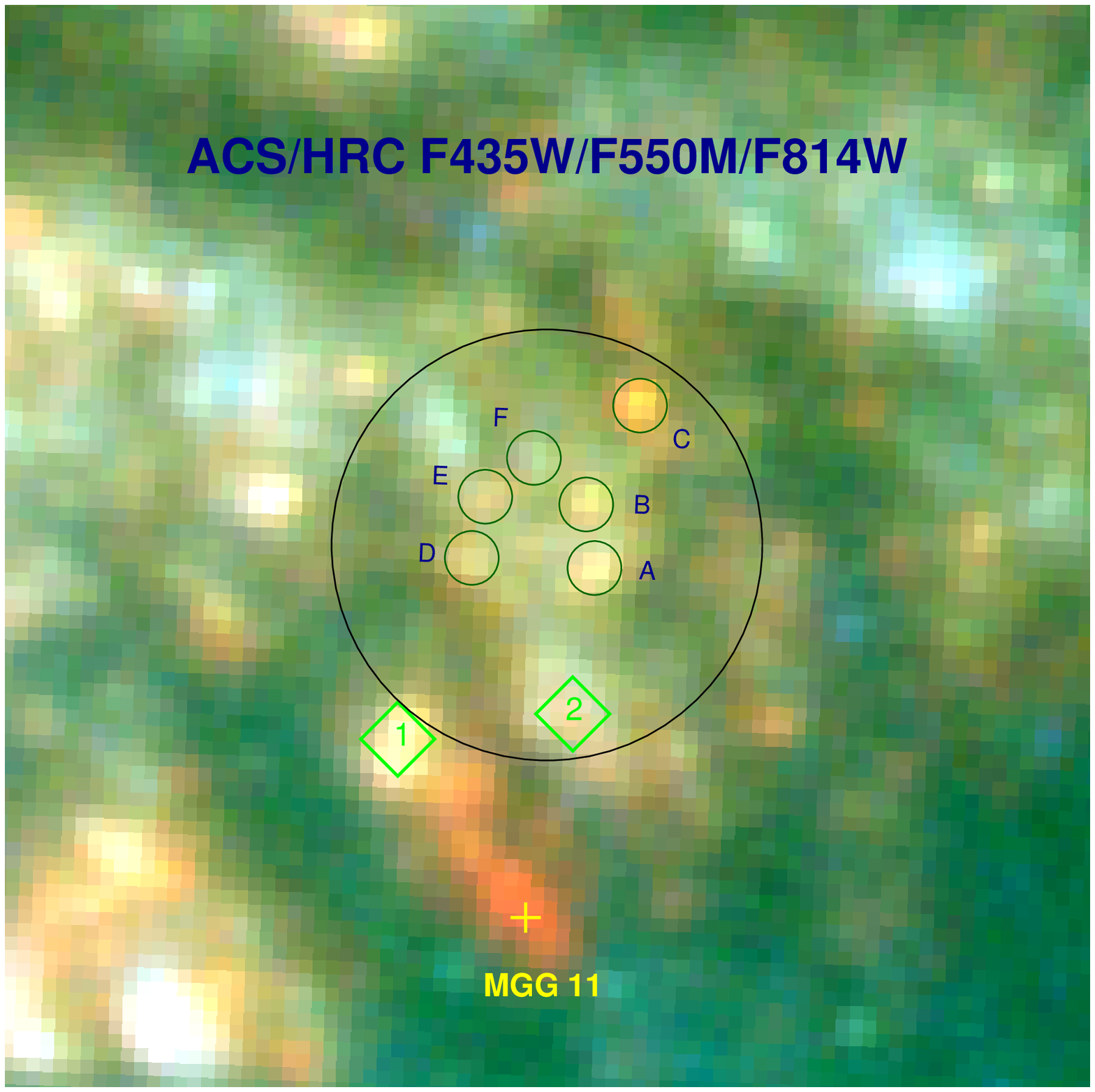}}
\drawline(-252.,152.)(-188.,187.)
\drawline(-252.,42.)(-188.,5.)
\caption{Left panel: The {\it HST} NICMOS image around M82 X-1. The $0.4''$ circle shows
the corrected X-ray position, and the box shows a region of $2''\times2''$.
The plus marks the position of the super star cluster MGG 11.
Right Panel: False-color ACS/HRC map around M81 X-1, using the F435W (blue), F550M (green), and F814W (red).
The six small circles, with radii of $0.05''$, represent six point sources located inside of the $0.4''$ circle.
The green diamonds show the two candidate counterparts identified by \citet{Gladstone2013},
both of which have been resolved into several sources due to the high resolution of ACS/HRC.}
\label{fig1}
\end{figure*}

To better resolve the objects in the crowded region around M82 X-1,
the observations from ACS/HRC, with a higher spatial resolution than that of ACS/WFC, are scrutinized.
As shown in Fig. \ref{fig1}, M82 X-1 is located beyond the super star cluster MGG 11, and
there are six point sources (A to F) located inside the $0.4''$ error circle,
which is the equivalent radius ($\sim~\sqrt{0.48''\times0.33''}$) of its error ellipse.
No other point sources are detected with IRAF/DAOFIND or DOLPHOT \footnote{http://americano.dolphinsim.com/dolphot/}.
Among the six objects,
source A lies closest to the regional center of M82 X-1 ($\Delta{r} \approx 0.096''$).
Source C and F is the reddest and bluest one, respectively.
The two optical counterpart candidates identified by \citet{Gladstone2013} are shown with green diamonds,
both of which are resolved into several sources by virtue of the high spatial resolution of ACS/HRC.
Although one of these is located inside the $0.4''$ circle, it is
quite extended, and is not considered as a counterpart.

\subsection{Photometry}
\label{phot.sec}

Aperture photometry for the six sources is performed with a small aperture radius ($0.05''$) to avoid contamination from nearby objects in this crowded region.
A small annulus immediately surrounding each source,
with inner radius as $0.06''$ and outer radius as $0.08''$,
is adopted to obtain the background intensity,
considering that these sources are superimposed onto a bright and variable background.
The photometry is performed using the {\sc DAOPHOT} package in IRAF on drizzled images produced from {\it HST}
standard pipeline calibration.
The {\sc WREGISTER} package in IRAF is used to align the frames in the same observation sequence.
For example, all the 16 WFC images in 2006 year are aligned to j9l024dtq\_drz.fits.
A bright nearby star is used to correct the positions of the six objects for observations in different sequences.
Although photometry with point-spread function (PSF) fitting is more robust than aperture
photometry in dense environments, the six objects are so faint that the flux profiles are poorly fit with the PSF.
The PSF-fitting package DOLPHOT is experimented to derive photometry,
but the results are poor and unacceptable. Therefore, only the results from {\sc DAOPHOT} are presented here.

The Vega magnitude zero points (ZPTs) are then adopted to compute the {\it HST} filter-dependent Vega magnitudes
for each source.
For ACS observations, the ZPTs referring to
an ``infinite'' radius are taken from \citet{Sirianni2005}. For WFC3 data, the ZPTs
corresponding to a $0.4''$ radius aperture are taken from the WFC3 Data Handbook.
We then derive the point-spread function models from Tiny Tim
\footnote{http://tinytim.stsci.edu/cgi-bin/tinytimweb.cgi.}, and
determine the aperture corrections from the $0.05''$ radius aperture
to the $0.5''$ (ACS) or $0.4''$ (WFC3) radius aperture using these models.
For ACS data, additional corrections from a $0.5''$ radius to an ``infinite'' radius
are taken \citep{Sirianni2005}.

The{\it HST} observations and computed magnitudes for the six sources
are listed in Table \ref{table2}.
The magnitude uncertainty only includes the error in the aperture sum,
  while the error due to background is not included,
  the latter of which is considered as a statistical effect.
  It should be noted that there are some variations in the photometry
  between different frames of the same exposure sequence
  (e.g., the four WFC/F435W images observed in 2006 year),
  and the variations can reach even one mag in some cases.
  In order to check the photometry accuracy, new photometry is performed
  on the {\it HST} observations with astrometric offsets corrected by TweakReg
  \footnote{http://ssb.stsci.edu/doc/stsci\_python\_dev/drizzlepac.doc/html/
tweakreg.html}.
  The test gives approximate photometry results, indicating
  the variations among these exposures are real.
  In addition, some photometry are performed for several small regions near these sources
  to check the background fluctuations.
  Similar variations between different frames of the same exposure sequence are found for these background regions,
  suggesting these variations are mostly caused by background fluctuations
  that would influence the photometry of faint objects,
  rather than being short-term variabilities of these point sources.
  This is not surprising, given the substantial and extensive diffuse emission in the 
  nuclear starburst region of M82 \citep{McCrady2003}.

\section{SED of the Counterparts}
\label{sed.sec}

\begin{figure*}[!htb]
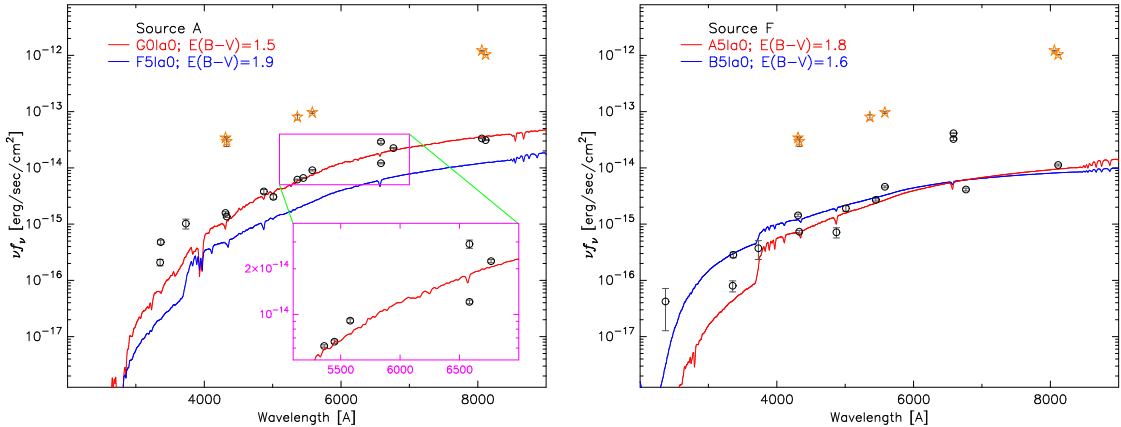

\figurenum{2}
\center
\subfigure{\includegraphics[width=0.4\textwidth]{object4.ps}}
\hspace{2 mm}
\subfigure{\includegraphics[width=0.4\textwidth]{objectF.ps}}
\caption{Left panel: The SED of source A. The black circles represent the apparent
magnitudes of source A, while the stars represent the magnitudes of a nearby super star
cluster MGG 6. A mean magnitude is plotted for observations with the same detector and filter.
The red and blue line represent the spectra of a supergiant G0 Ia0 ($M_V=-8.9$) and F5 Ia0 ($M_V=-9.0$),
with different extinction correction, respectively.
Right panel: The SED of source F. The black circles represent the apparent
magnitudes of source F, and the stars indicate the SED of MGG 6.
The red and blue line represent the spectra of a supergiant A5 Ia0 ($M_V=-8.8$) and B5 Ia0 ($M_V=-8.4$),
with different extinction correction, respectively.}
\label{fig2}
\end{figure*}

Spectral energy distributions for these six objects are constructed from {\it HST} photometry,
and two examples (object A and F) are plotted in Fig. \ref{fig2}.
The SED of a nearby super star cluster MGG 6 is overplotted for comparison, which shows
a much steeper trend towards the red band due to numbers of dwarf stars;
the red band of object A and F, like stellar objects, is much flatter.
The SED of MGG 6 displays much brighter magnitudes (${\rm F814W \approx 17}$ mag) and redder colors
than all the six objects.
Quantitatively, the mean color ${\rm F814W_{HRC}-F550M_{HRC}}$ for the six objects is 2.2 mag,
while ${\rm F814W_{HRC}-F550M_{HRC}} = 3.5$ mag for MGG 6.
Although no exact intrinsic color can be obtained without accurate extinction,
such notable differences suggest that these six objects are likely stellar objects instead of super star clusters.

To classify the counterparts, SED templates are calculated with the SYNPHOT tool {\sc CALCSPEC}
using the CK04 standard stellar spectra \citep{Castelli2004} with solar metallicity.
All SED templates are normalized to a $V$-band magnitude of zero for simple comparisons.
The standard stars in the CK04 models, whose absolute magnitudes are taken from
\citet{Schmidt-Kaler1982} and \citet{Martins2005}, are placed at the distance of M82
\citep[3.63 Mpc; ][]{Freedman1994}.
Assuming Galactic relation $N_{\rm H} = 5.8\times10^{21}E(B-V)$ \citep{Bohlin1978},
the extinction toward M82 X-1 is calculated as $E(B-V) \simeq 1.9$,
with $N_{\rm H} \simeq 1.1\times10^{22}~{\rm cm}^{-2}$ inferred from the X-ray spectrum of the ULX \citep{Feng2010}.
Note that, the X-ray based $N_{\rm H}$ absorption may include contribution from the accretion disk itself, which may have little effect on the optical observation
of the secondary, thus a lower extinction is possible for the secondary.
In the following analysis, variable $E(B-V)$ values ranging from 1.0 to 1.9 are adopted in comparing the observed SEDs and the templates (from O to M).

The red part ($\lambda >$ 4000 {\AA}) of the observed SEDs and the templates are used for the comparison, since
the blue emission of an X-ray binary may be contaminated by emission from the accretion disk and
reprocessed emission from an X-ray illuminated accretion disk or stellar surface \citep{Tao2011},
the last of which could brighten the system by up to $\sim 5$ mag \citep{Copperwheat2007}.
With simple comparisons,
the SEDs of sources A, B, C, D, E, and F match well with those of supergiants, with types as
F5-G0, A0-F5, K5-M0, A5-F5, F0-F5, and B5-G0, respectively.
Here no exact fitting is done in the comparison, because the extinction can vary in a wide range.
It should be noted that the wide range of the spectral type for each source
is due to the wide range of the extinction.

\section{Discussion}
\label{discuss.sec}

In this work, we have carried out careful relative astrometry and identified six objects in the immediate region
around M82 X-1, and also estimated their types from their observed SEDs constructed from {\it HST} photometry.

Intriguingly,  both source A (in Data Set j8mx19010 and j9fb09030)
and F (in all observations) show excessive fluxes relative to the optical continuum in the narrow $H_{\alpha}$ filter F658N.
Note that the background intensity has been carefully subtracted, and
there is no F658N flux excess in the SEDs of other nearby sources (B, C, D, and E).
Generally, $H_{\alpha}$ emission can come from H II regions, planetary nebulae, or accretion disks.
Both H II regions and planetary nebulae produce significant S II (6731 {\AA}) emission in the narrow S II filter F673N along with $H_{\alpha}$ emission,
yet no excessive S II emission is seen in the F673N observations for any of the objects.
This suggests that the $H_{\alpha}$ emission of source A and F does not come from H II regions or planetary nebulae,
but is likely produced by an accretion disk, hence suggesting that one of the two sources is the optical counterpart of M82 X-1.

For source A,
the blue part of the SED is brighter than that of a late-type companion star, suggesting
that the blue component may come from combined emission of the companion star and the accretion disk,
while the red component may come mainly from the companion star.
Except for source A, the other five objects do not show clear blue excess in the SEDs.
Source F is the bluest one among the six objects, and is hardly seen in the F814W images.
This may suggest the blue emission comes from an accretion disk,
as some studies \citep{Pakull2006, Tao2011} have claimed
the dominant component of optical emission
in most ULXs is produced through X-ray irradiation of the outer accretion disk.
Sources F, D and E exhibit variability in multiple observations, however,
the crowded environments and bright background restrain us from confirming whether the long-term flux variability
is real. 
In addition, sources D and E seem a little more extended than other four objects,
but the SEDs indicate they could not be star clusters or peaks in the diffuse emission (no feature of H II region).

The scenario of a supergiant companion is consistent with the orbital period of 62 days of M82 X-1,
and the short phase with Roche lobe overflowing indicates
M82 X-1 is in a brief and unusual period of its evolution \citep{Kaaret2006a}.
Further deep spatially resolved spectroscopic observation would provide more information on
the two candidates, and unveil the nature of M82 X-1.

\acknowledgements
\begin{acknowledgements}
We especially thank the anonymous referee for his/her thorough report and helpful comments
and suggestions that have significantly improved the paper.
This work is based on observations made with the NASA/ESA {\it Hubble Space Telescope},
obtained from the Mikulski Archive for Space Telescopes.
Some of the data presented in this paper were obtained from the {\it Chandra} Data Archive and SDSS-III.
The authors acknowledge support from the National Science Foundation of China under grants
NSFC-11273028 and NSFC-11333004, and support from the National Astronomical Observatories,
Chinese Academy of Sciences under the Young Researcher Grant.

\end{acknowledgements}

\begin{table}
\renewcommand{\arraystretch}{1.2}
\begin{center}
\caption[]{Positions of Sources for {\it Chandra}-SDSS and SDSS-{\it HST} Astrometric Registration.}
\label{table1}\small
\begin{tabular}{ccccc}
\tableline
\tableline
\multicolumn{5}{c}{Bright {\it Chandra} ACIS X-ray Sources identified in SDSS Observations}\\
\hline
{\it Chandra} R.A. &    {\it Chandra} Decl. &   SDSS R.A. &       SDSS Decl. &      Displacement ($''$)\\
\hline
9:56:58.658 &     +69:38:52.12 &    9:56:58.670 &     +69:38:52.53 &    0.423       \\
9:55:43.066 &     +69:34:55.20 &    9:55:43.018 &     +69:34:54.82 &    0.461       \\
9:55:05.222 &     +69:44:42.47 &    9:55:05.182 &     +69:44:42.48 &    0.220       \\
9:55:14.585 &     +69:47:35.59 &    9:55:14.530 &     +69:47:35.74 &    0.331       \\
9:55:34.560 &     +69:38:23.32 &    9:55:34.567 &     +69:38:24.00 &    0.681       \\
9:55:58.613 &     +69:40:47.17 &    9:55:58.553 &     +69:40:47.25 &    0.331       \\
\hline
\multicolumn{5}{c}{Bright SDSS Sources identified in {\it HST} ACS/WFC Observations}\\
\hline
SDSS R.A. &       SDSS Decl. &      {\it HST} R.A. &        {\it HST} Decl. &       Displacement ($''$)\\
\hline
9:55:47.134 &     +69:40:41.87 &    9:55:47.002 &     +69:40:41.71 &    0.727       \\
9:55:41.806 &     +69:41:15.41 &    9:55:41.664 &     +69:41:15.26 &    0.777       \\
9:55:46.805 &     +69:40:38.09 &    9:55:46.656 &     +69:40:37.99 &    0.806       \\
9:55:55.711 &     +69:41:05.42 &    9:55:55.546 &     +69:41:05.29 &    0.901       \\
\tableline
\end{tabular}
\end{center}
\end{table}

\begin{table}
\renewcommand{\arraystretch}{1.2}
\begin{center}
\caption[]{{\it HST} Observations and Magnitudes of Several Sources in the Region of M82 X-1.}
\label{table2}\small
\tabcolsep=1.pt
\begin{tabular}{cccccccccc}
\tableline
\tableline
&              &              &               &           A &               B &               C &               D &               E &               F                 \\
Date &         Data Set &     Instrument/Filter &   Exp(s) &$m_{\rm Filter}$ &$m_{\rm Filter}$ &$m_{\rm Filter}$ &$m_{\rm Filter}$ &$m_{\rm Filter}$ &$m_{\rm Filter}$   \\
\hline
2004-02-09 &   j8mx19010 &    ACS/WFC/F658N &     700 &          21.523$\pm$0.048 &22.139$\pm$0.063 &23.318$\pm$0.109 &24.480$\pm$0.186 &22.808$\pm$0.086 &20.954$\pm$0.037   \\
2004-02-09 &   j8mx19e9q &    ACS/WFC/F814W &     120 &          20.812$\pm$0.020 &21.161$\pm$0.023 &20.085$\pm$0.014 &22.255$\pm$0.039 &22.717$\pm$0.048 &...  \\
2005-12-08 &   j9fb09010 &    ACS/HRC/F330W &     3896 &         25.890$\pm$0.091 &27.136$\pm$0.161 &... &26.234$\pm$0.106 &... &26.431$\pm$0.116   \\
2005-12-08 &   j9fb09030 &    ACS/HRC/F658N &     400 &          21.506$\pm$0.066 &... &21.410$\pm$0.063 &22.673$\pm$0.112 &22.404$\pm$0.099 &21.383$\pm$0.062   \\
2006-01-31 &   j9fb57030 &    ACS/HRC/F435W &     1132 &         25.593$\pm$0.052 &25.510$\pm$0.050 &28.171$\pm$0.171 &26.308$\pm$0.072 &26.677$\pm$0.086 &25.700$\pm$0.055   \\
2006-01-31 &   j9fb57010 &    ACS/HRC/F550M &     840 &          23.276$\pm$0.031 &23.495$\pm$0.034 &22.977$\pm$0.027 &23.878$\pm$0.041 &24.450$\pm$0.053 &24.012$\pm$0.043   \\
2006-01-31 &   j9fb57020 &    ACS/HRC/F814W &     140 &          21.151$\pm$0.025 &21.448$\pm$0.029 &20.214$\pm$0.016 &21.864$\pm$0.035 &21.948$\pm$0.036 &22.256$\pm$0.042   \\
2006-03-27 &   j9l021d6q &    ACS/WFC/F435W &     450 &          25.660$\pm$0.081 &25.340$\pm$0.070 &26.793$\pm$0.137 &... &26.189$\pm$0.104 &26.571$\pm$0.124   \\
2006-03-27 &   j9l022deq &    ACS/WFC/F435W &     450 &          25.696$\pm$0.083 &26.375$\pm$0.113 &26.572$\pm$0.124 &... &27.045$\pm$0.154 &26.050$\pm$0.097   \\
2006-03-27 &   j9l023dmq &    ACS/WFC/F435W &     450 &          25.886$\pm$0.090 &26.461$\pm$0.118 &26.800$\pm$0.137 &... &27.045$\pm$0.154 &27.305$\pm$0.174   \\
2006-03-27 &   j9l024duq &    ACS/WFC/F435W &     450 &          25.815$\pm$0.087 &25.438$\pm$0.073 &26.938$\pm$0.147 &... &26.231$\pm$0.106 &26.209$\pm$0.105   \\
2006-03-27 &   j9l021d8q &    ACS/WFC/F555W &     340 &          23.578$\pm$0.036 &23.891$\pm$0.042 &23.746$\pm$0.039 &24.592$\pm$0.057 &25.603$\pm$0.091 &...  \\
2006-03-27 &   j9l022dgq &    ACS/WFC/F555W &     340 &          23.792$\pm$0.040 &23.913$\pm$0.042 &23.696$\pm$0.038 &25.414$\pm$0.084 &25.789$\pm$0.099 &...  \\
2006-03-27 &   j9l023doq &    ACS/WFC/F555W &     340 &          24.077$\pm$0.045 &24.112$\pm$0.046 &24.060$\pm$0.045 &25.245$\pm$0.077 &25.366$\pm$0.082 &...  \\
2006-03-27 &   j9l024dwq &    ACS/WFC/F555W &     340 &          23.779$\pm$0.039 &23.865$\pm$0.041 &23.800$\pm$0.040 &24.888$\pm$0.066 &25.565$\pm$0.090 &...  \\
2006-03-27 &   j9l021daq &    ACS/WFC/F658N &     1100 &         22.492$\pm$0.059 &23.114$\pm$0.079 &22.620$\pm$0.063 &23.698$\pm$0.103 &22.345$\pm$0.055 &21.114$\pm$0.031   \\
2006-03-27 &   j9l022diq &    ACS/WFC/F658N &     1100 &         23.812$\pm$0.108 &22.231$\pm$0.052 &21.083$\pm$0.031 &... &21.851$\pm$0.044 &21.065$\pm$0.031   \\
2006-03-27 &   j9l023dqq &    ACS/WFC/F658N &     1100 &         23.649$\pm$0.101 &22.487$\pm$0.059 &21.020$\pm$0.030 &... &22.688$\pm$0.065 &21.315$\pm$0.034   \\
2006-03-27 &   j9l024dyq &    ACS/WFC/F658N &     1100 &         22.329$\pm$0.055 &22.938$\pm$0.073 &20.959$\pm$0.029 &... &22.425$\pm$0.057 &21.129$\pm$0.032   \\
2006-03-27 &   j9l021d5q &    ACS/WFC/F814W &     175 &          21.068$\pm$0.018 &21.489$\pm$0.022 &20.151$\pm$0.012 &21.528$\pm$0.023 &22.824$\pm$0.041 &...  \\
2006-03-27 &   j9l022ddq &    ACS/WFC/F814W &     175 &          21.438$\pm$0.022 &21.557$\pm$0.023 &20.456$\pm$0.014 &22.066$\pm$0.029 &22.050$\pm$0.029 &...  \\
2006-03-27 &   j9l023dlq &    ACS/WFC/F814W &     175 &          21.254$\pm$0.020 &21.665$\pm$0.024 &20.569$\pm$0.015 &21.952$\pm$0.028 &22.428$\pm$0.035 &...  \\
2006-03-27 &   j9l024dtq &    ACS/WFC/F814W &     175 &          20.882$\pm$0.017 &21.190$\pm$0.020 &20.292$\pm$0.013 &21.339$\pm$0.021 &22.166$\pm$0.031 &...  \\
2009-11-17 &   ib6w81060 &    WFC3/UVIS/F225W &   1070 &         ... &27.031$\pm$0.416 &27.218$\pm$0.453 &26.384$\pm$0.309 &... &...  \\
2009-11-17 &   ib6w81050 &    WFC3/UVIS/F336W &   1050 &         26.861$\pm$0.228 &27.352$\pm$0.286 &... &27.785$\pm$0.349 &... &27.552$\pm$0.314   \\
2009-11-15 &   ib6w83020 &    WFC3/UVIS/F373N &   2850 &         25.194$\pm$0.196 &24.981$\pm$0.178 &28.094$\pm$0.746 &25.774$\pm$0.257 &... &26.383$\pm$0.340   \\
2009-11-17 &   ib6w82030 &    WFC3/UVIS/F487N &   2455 &         24.142$\pm$0.087 &25.558$\pm$0.167 &24.807$\pm$0.118 &24.652$\pm$0.110 &25.128$\pm$0.137 &25.989$\pm$0.203   \\
2009-11-17 &   ib6w82020 &    WFC3/UVIS/F502N &   2465 &         24.637$\pm$0.093 &... &25.601$\pm$0.144 &... &... &25.139$\pm$0.117   \\
2009-11-17 &   ib6w81040 &    WFC3/UVIS/F547M &   360 &          24.072$\pm$0.065 &23.862$\pm$0.059 &24.554$\pm$0.081 &24.220$\pm$0.069 &25.005$\pm$0.100 &24.381$\pm$0.075   \\
2009-11-15 &   ib6w83030 &    WFC3/UVIS/F673N &   2760 &         21.814$\pm$0.026 &22.068$\pm$0.029 &21.310$\pm$0.020 &23.750$\pm$0.063 &24.034$\pm$0.071 &23.650$\pm$0.060   \\
2010-01-01 &   ib6wr9030 &    WFC3/UVIS/F225W &   1665 &         ... &... &... &... &... &28.206$\pm$0.575   \\
2010-01-01 &   ib6wr9020 &    WFC3/UVIS/F336W &   1620 &         26.509$\pm$0.157 &... &26.989$\pm$0.195 &28.275$\pm$0.353 &26.280$\pm$0.141 &27.828$\pm$0.287   \\
2010-01-01 &   ib6wr9010 &    WFC3/UVIS/F547M &   1070 &         23.244$\pm$0.026 &23.615$\pm$0.030 &24.050$\pm$0.037 &24.884$\pm$0.055 &25.251$\pm$0.065 &24.757$\pm$0.052   \\
\tableline
\end{tabular}
\end{center}
\end{table}

\end{document}